# First-principles study of the effects of electron-phonon coupling on the thermoelectric properties: a case study of SiGe compound

D. D. Fan, H. J. Liu[*], L. Cheng, J. H. Liang, P. H. Jiang

*Key Laboratory of Artificial Micro- and Nano-Structures of Ministry of Education and School of Physics and Technology, Wuhan University, Wuhan 430072, China*

It is generally assumed in the thermoelectric community that the lattice thermal conductivity of a given material is independent of the electronic properties. This perspective is however questionable since the electron-phonon coupling could have certain effects on both the carrier and phonon transport, which in turn will affect the thermoelectric properties. Using SiGe compound as a prototypical example, we give an accurate prediction of the carrier relaxation time by considering scattering from all the phonon modes, as opposed to the simple deformation potential theory. It is found that the carrier relaxation time does not change much with the concentration, which is however not the case for the phonon transport where the lattice thermal conductivity can be significantly reduced by electron-phonon coupling at higher carrier concentration. As a consequence, the figure-of-merit of SiGe compound is obviously enhanced at optimized carrier concentration, and becomes more pronounced at elevated temperature.

## 1. Introduction

The strive for efficient thermoelectric generators has attracted great attention due to the potential application in refrigeration and power generation [1]. The performance of a thermoelectric material is usually characterized by the figure-of-merit $ZT = S^2\sigma T/\kappa$, where $S$, $\sigma$, $T$, $\kappa$ are the Seebeck coefficient, the electrical conductivity, the absolute temperature, and the thermal conductivity (including the electronic part $\kappa_e$ and the phonon part $\kappa_p$), respectively. High performance thermoelectric materials require large power factor ($S^2\sigma$) and/or low thermal conductivity. The past two decades have witnessed the rapid development of thermoelectrics, where various innovative strategies have been applied to improve the efficiency of thermoelectric materials, such as band convergence [2, 3], resonance energy level [4], nanostructuring [5], and all-scale hierarchical architectures [6].

---

[*] Author to whom correspondence should be addressed. Electronic mail: phlhj@whu.edu.cn

It is well known that the electron-phonon coupling (EPC) plays an important role in governing the electronic transport properties, and the three-phonon process is the main scattering mechanism for the phonon transport in a crystal. For a long time, it is generally assumed that the lattice thermal conductivity of a given material is independent of the electronic properties. However, since the EPC has a significant effect on the electronic transport properties, it should also have certain effects on the phonon transport. Fortunately, there are several experimental studies that have captured the relevance of the lattice thermal conductivity with the electronic properties. For example, Bentien *et al.* [7] found a fundamental difference in $\kappa_p$ between *n*- and *p*-type samples of the thermoelectric material $Ba_8Ga_{16}Ge_{30}$, and proposed that the phonon-carrier scattering must be considered in order to explain $\kappa_p$ at low temperature. In a further study, they [8] showed that $\kappa_p$ is intimately linked to the charge carriers, and the glass-like $\kappa_p$ was considered as a consequence of strong phonon-carrier coupling. In addition, May *et al.* [9] investigated the electron and phonon scattering in the high temperature thermoelectric material $La_3Te_{4-z}M_z$ (M = Sb, Bi) and found that $\kappa_p$ increases with decreasing $\sigma$, which was considered as the possible evidence of the EPC. In the theoretical aspect, particularly worth mentioning is that Liao *et al.* [10] revealed an unexpectedly significant reduction (45%) of the lattice thermal conductivity of bulk silicon by considering the EPC at a high carrier concentration ($10^{21}$ cm$^{-3}$). Indeed, experimental study of Zhu *et al.* [11] found a large reduction (36%) of $\kappa_p$ in heavily doped silicon ($Si_{0.94}P_{0.06}$). These pioneering works have stimulated a lot of subsequent studies to investigate the effects of EPC on the lattice thermal conductivity in metals and semiconductors [12, 13, 14, 15]. The common wisdom suggested that the EPC would be very important to the thermal transport at high carrier concentration. For example, a maximum drops by 57% was predicted in the lattice thermal conductivity of 3C-SiC with a carrier concentration of $10^{21}$ cm$^{-3}$ [15]. Moreover, since the electronic transport coefficients in the $ZT$ are usually coupled with each other, one of the most effective schemes is to minimize $\kappa_p$ [6] to obtain better thermoelectric performance. It is thus of vital importance to investigate the effect of EPC on the thermoelectric properties, especially at higher carrier concentration.

In this work, using SiGe compound as a prototypical example, we present a complete study of the effects of EPC on its electronic and phonon transport properties. Note that SiGe systems are known as traditional high temperature thermoelectric materials [16]. Intriguingly, it is found that during the EPC process, the carrier relaxation time is somehow insensitive to the carrier concentration while the phonon relaxation time shows a marked dependence. Considering both the intrinsic phonon-phonon scattering and that from EPC, we demonstrate that the lattice thermal conductivity of SiGe compound can be reduced significantly with increasing carrier concentration. As a result, a 24% increase in the $ZT$ value can be found at 1200 K.

## 2. Computational method

The electronic properties of SiGe compound are investigated within the framework of density functional theory (DFT), as implemented in the so-called Quantum ESPRESSO package [17]. The norm-conserving scalar-relativistic pseudopotential [18] is adopted to describe the core-valence interaction, and the exchange-correlation functional is in the form of Perdew-Bruke-Ernzerhof (PBE) with the generalized gradient approximation (GGA) [19]. The density functional perturbation theory (DFPT) [20] is employed to calculate the phonon dispersion relations. Meanwhile, the electronic and phonon transport properties are investigated by using the Boltzmann transport theory [21]. The key point for accurate calculations of the transport coefficients is appropriate treatment of the scattering mechanisms. It is well known that the EPC plays an important role in governing the electronic transport. Earlier attempts in addressing this issue adopted the deformation potential theory [22], which considers only the electron-acoustic phonon interactions. The electron relaxation time thus obtained is assumed to be a constant [23, 24] which is generally overestimated [25, 26]. Instead, we obtain the $k$-resolved relaxation time $\tau_{n\mathbf{k}}(\mu,T)$ from the imaginary part of the electron self-energy by a complete EPC calculation [27]

$$\frac{1}{\tau_{n\mathbf{k}}(\mu,T)} = \frac{2\pi}{\hbar} \sum_{m,\mathbf{q}v} |g_{mn}^{v}(\mathbf{k},\mathbf{q})|^2 \left\{ \left[ n(\omega_{\mathbf{q}v},T) + f(\varepsilon_{m\mathbf{k}+\mathbf{q}},\mu,T) \right] \delta(\varepsilon_{n\mathbf{k}} - \varepsilon_{m\mathbf{k}+\mathbf{q}} + \hbar\omega_{\mathbf{q}v}) \right. \\ \left. + \left[ n(\omega_{\mathbf{q}v},T) + 1 - f(\varepsilon_{m\mathbf{k}+\mathbf{q}},\mu,T) \right] \delta(\varepsilon_{n\mathbf{k}} - \varepsilon_{m\mathbf{k}+\mathbf{q}} - \hbar\omega_{\mathbf{q}v}) \right\}, \quad (1)$$

where the key variable is the EPC matrix element $g_{mn}^{v}(\mathbf{k},\mathbf{q})$, which characterizes the strength of the EPC and is given by

$$g_{mn}^{v}(\mathbf{k},\mathbf{q}) = \langle \psi_{m\mathbf{k}+\mathbf{q}} | \partial_{\mathbf{q}v} V | \psi_{n\mathbf{k}} \rangle. \quad (2)$$

Here $\psi_{m\mathbf{k}}$ is the wavefunciton of band $m$ and wavevector $k$, $\varepsilon_{m\mathbf{k}}$ is the corresponding energy eigenvalue, and $\partial_{\mathbf{q}v} V$ is the derivative of the self-consistent potential associated with a phonon mode having frequency $\omega_{\mathbf{q}v}$ at wavevector $q$ and polarization $v$. $n(\omega_{\mathbf{q}v},T)$ and $f(\varepsilon_{m\mathbf{k}},\mu,T)$ are respectively the Bose-Einstein and Fermi-Dirac distribution function, where the temperature $T$ and chemical potential $\mu$ (corresponds to carrier concentration) enter into the carrier relaxation time. The influence of the carrier concentration on the relaxation time will be discussed in details later. Combined with the Boltzmann transport equation, such scheme has been proved to yield good agreement with experiments [28, 29, 30]. For the heat transport, the total phonon scattering rate $1/\tau_{tot,v\mathbf{q}}(\mu,T)$ is evaluated by including both the intrinsic phonon-phonon interactions and EPC according to the Mattiessen's rule [21]

$$1/\tau_{tot,v\mathbf{q}}(\mu,T) = 1/\tau_{p-p,v\mathbf{q}}(T) + 1/\tau_{e-p,v\mathbf{q}}(\mu,T), \quad (3)$$

where $1/\tau_{e-p,v\mathbf{q}}(\mu,T)$ is the phonon scattering rate in the EPC process obtained from the imaginary part of the phonon self-energy [27]

$$1/\tau_{v\mathbf{q},ep}(\mu,T) = \frac{2\pi}{\hbar} \sum_{mn,\mathbf{k}} |g_{mn}^{v}(\mathbf{k},\mathbf{q})|^2 \left[ f(\varepsilon_{n\mathbf{k}},\mu,T) - f(\varepsilon_{m\mathbf{k}+\mathbf{q}},\mu,T) \right] \\ \times \delta(\varepsilon_{n\mathbf{k}} - \varepsilon_{m\mathbf{k}+\mathbf{q}} - \hbar\omega_{v\mathbf{q}}), \quad (4)$$

and the intrinsic phonon relaxation time $\tau_{p-p,\nu\mathbf{q}}(T)$ is calculated by considering the three-phonon process as well as the effect of isotopic disorder [31]. For the calculation of intrinsic phonon relaxation time, the second- and third-order interatomic force constants (IFCs) are both calculated by the finite difference approach using a 5×5×5 supercell. The fourth nearest neighbors are included for the third-order interactions, which corresponds to a cutoff distance of 5.9 Å. To achieve converged results, very dense sampling points are necessary when doing the Brillouin zone integrations, especially for the calculations of electronic transport coefficients [28, 29, 30, 32]. To find balance between convergence and computational efficiency (see Figure S1 of the Supporting Information), we adopt the Wannier interpolation scheme as implemented in the EPW [17] package. That is, the band structure, the phonon spectrum and the EPC matrix elements are first calculated on the coarse grids of 10×10×10 $k$-points and 5×5×5 $q$-points, and are then interpolated to the dense 100×100×100 $k$-points and 25×25×25 $q$-points.

## 3. Results and discussion

The SiGe compound crystallizes in the zinc blende structure with space group $F\bar{4}3m$. The optimized lattice constant is 5.60 Å. Figure 1(a) displays the band structure of SiGe compound calculated by DFT and the Wannier interpolation technique [33]. The results agree very well with each other, indicating the reliability of Wannierization. The band gap predicted by PBE functional is 0.61 eV, which is increased to 1.16 eV by using the hybrid density functional [34]. Such a correction is very important when performing the electronic transport calculation, especially for the case of high temperature thermoelectric material such as SiGe systems (see Figure S2 of the Supporting Information). It is noted that the conduction band minimum (CBM) located at $\Delta$ point obviously has a larger effective mass than that of the valence band maximum (VBM), suggesting a higher $n$-type Seebeck coefficient. We thus focus on the $n$-type system in the following, where the carrier concentration can be derived from the corresponding chemical potential, as illustrated by the dashed lines in Fig. 1(a). The phonon spectrum of SiGe compound is shown in Fig. 1(b), where we find no imaginary frequency indicating the dynamic stability of the system. Besides, we discover a large acoustic-optical gap of ~ 100 cm$^{-1}$, leading to a significant reduction in the phonon scattering rate due to the prohibited three-phonon process [35]. Moreover, the optical branches of SiGe compound exhibit less dispersions, which usually have small contribution to the heat transport [36]. All these observations indicate that the lattice thermal conductivity of SiGe compound is largely determined by the low frequency acoustic phonon modes.

To have a deep insight in the transport properties of the SiGe compound, we plot in Figure 2 the room temperature relaxation time of both electron and phonon, where the results at two typical carrier concentrations of $10^{19}$ and $10^{21}$ cm$^{-3}$ are compared. We see from Fig. 2(a) that the electron relaxation time is large around the band edges. Since the phonon energy is almost negligible compared with that of electron, the latter can thus be approximately considered to be scattered to a constant energy surface in the EPC process. In another word, the scattering channels for the electrons are strongly limited near CBM [37]. Besides, we see that the electron relaxation time calculated at $10^{19}$ and $10^{21}$ cm$^{-3}$ coincides with each other when the energy is far away from the chemical potential. This result can be explained by the fact that the occupation number of the electron states far below (above) the chemical potential is nearly one (zero). The Fermi-Dirac function in Eq. (1) can be thus replaced by 1 or 0 such that the electron relaxation time does not have obvious carrier concentration dependence. It is interesting to find that the maximum relaxation time (at CBM) only reduces by half when the carrier concentration increases by 100 times. Such observation can be understood by referring to the two terms with $\delta$ function in Eq. (1), which have opposite effects on the electron relaxation time. The physical meaning is that the first term describes the process that an electron in the state ***k*** absorbs a phonon in the state ***q*** and is scattered to the state ***k***+***q***, while the second term represents the reversed process. Compared with that of electron, the energy of phonon can be ignored thus the electron relaxation time in Eq. (1) can be simplified as

$$\frac{1}{\tau_{n\mathbf{k}}(\mu,T)} = \frac{2\pi}{\hbar} \sum_{m,\mathbf{q}\nu} \left|g_{mn}^{\nu}(\mathbf{k},\mathbf{q})\right|^2 \left[2n(\omega_{\mathbf{q}\nu},T)+1\right]\delta(\varepsilon_{n\mathbf{k}} - \varepsilon_{m\mathbf{k}+\mathbf{q}}), \tag{5}$$

which suggests the weak dependence of the electron relaxation time on the carrier concentration. As a consequence, the electronic transport coefficients thus calculated are less sensitive to the electron relaxation time obtained at different carrier concentration, which is more pronounced at elevated temperature (see Fig. S3 of the Supporting Information). On the other hand, to fully understand the effect of EPC on the lattice thermal conductivity, we compare in Fig. 2(b) the phonon relaxation time originated from the EPC ($\tau_{e-p}$) and intrinsic phonon-phonon scattering ($\tau_{p-p}$), where results for two typical carrier concentration of $10^{19}$ and $10^{21}$ cm$^{-3}$ are both shown. In contrast to the weak carrier concentration dependence of the electron relaxation time, we see that $\tau_{e-p}$ is reduced by at least two orders of magnitude when the carrier concentration increases by 100 times. This is reasonable since any phonons can be scattered by electrons such that the phonon scattering rate is proportional to the number of electrons in the EPC process. At a higher carrier concentration of $10^{21}$ cm$^{-3}$, we find that the phonon relaxation time from EPC (red points) overlaps with that of intrinsic phonon scattering (blue points) in the range of $10^{-9} \sim 10^{-11}$ s. In particular, the phonon relaxation time from EPC becomes smaller than that from intrinsic phonon-phonon interaction within the frequency of 150 cm$^{-1}$, where 96% of the lattice thermal conductivity is contributed by the phonons in this range (see the inset). According to the Mattiessen's rule given in Eq. (3), we believe that the EPC should have significant effect on the lattice thermal conductivity, as shown in the following.

Figure 3(a) plots the room temperature lattice thermal conductivity ($\kappa_p$) of SiGe compound calculated at a series carrier concentration of $10^{18}$, $10^{19}$, $10^{20}$ and $10^{21}$ cm$^{-3}$ (blue circles), where the effect of EPC is explicitly taken into account (also see Table I in the Supporting Information for those at elevated temperature). For comparison, the electronic thermal conductivity ($\kappa_e$), the intrinsic lattice thermal conductivity ($\kappa_p^0$), and the total thermal conductivity with and without EPC correction ($\kappa_{tot}$ and $\kappa_{tot}^0$) are all shown. We see that $\kappa_p$ decreases at a faster rate when the carrier concentration becomes larger, which is consistent with those reported previously [10, 14]. It is interesting to find that the calculated lattice thermal conductivity can be fitted by

$$\kappa_p = \kappa_p^0 - 6.70 \times 10^{-12} e^{\log_{10}(n)/0.71}, \qquad (6)$$

where $n$ is the carrier concentration. Moreover, we find that the total thermal conductivity without EPC correction ($\kappa_{tot}^0$) increases with the carrier concentration due to the increase of the electronic thermal conductivity. On the contrary, the EPC corrected total thermal conductivity ($\kappa_{tot}$) decreases with carrier concentration since the lattice thermal conductivity decreases much faster than the increase of the electronic part. The difference between $\kappa_{tot}$ and $\kappa_{tot}^0$ becomes more pronounced at higher carrier concentration, which suggests the vital role of EPC on the thermoelectric performance of SiGe compound. In Fig. 3(b), we plot the room temperature *ZT* value as a function of carrier concentration. It is clear that considering EPC leads to enhanced *ZT* when the carrier concentration is larger than $10^{19}$ cm$^{-3}$, which just covers the optimized carrier concentration range for most thermoelectric materials [38, 39, 40, 41]. As SiGe systems are typical high temperature thermoelectric materials, in Fig. 3(c), we also show the thermal conductivity at 1200 K as a function of carrier concentration. At higher carrier concentration, we see that the electronic thermal conductivity (black solid line) competes with the lattice part (blue solid line), and becomes identical at a certain carrier concentration of $\sim 2 \times 10^{21}$ cm$^{-3}$. Before the crossing point, the lattice thermal conductivity is relatively higher so that the EPC correction could have a larger effect on the *ZT*. As carrier concentration is increased, the electronic thermal conductivity becomes important so that the EPC correction on the *ZT* becomes less pronounced. Such competition between the lattice thermal conductivity and the electronic thermal conductivity governs the EPC correction on the *ZT* value, as shown in Fig. 3(d). At lower carrier concentration (< $10^{19}$ cm$^{-3}$), the effect of EPC on the lattice thermal conductivity is quite small and thus the correction of *ZT* is negligible. With the increase of carrier concentration, the effect of EPC becomes more and more obvious, and there is a significant increase of the *ZT* value. For example, the *ZT* is increased by 24% at optimized concentration of $3\times10^{20}$ cm$^{-3}$, which emphasizes the vital importance of the EPC for correctly predicting the thermoelectric performance at relatively higher carrier concentration. It should be mentioned that the experimentally measured *n*-type *ZT* of SiGe alloy is about 1.4 [42, 43] around 1200 K, which is significantly larger than our calculated value of about 0.2. Such apparent discrepancy mainly comes from the fact that the experimental samples are usually in the form of alloy with disordered Si and Ge, while our studied SiGe compound is a highly ordered system. As a consequence, the calculated lattice thermal conductivity should be much higher than the experimentally reported values.

## 4. Conclusions

Using SiGe compound as a prototype, we present a first-principles study of the effect of EPC on the thermoelectric transport properties. It is found that the electron relaxation time is less affected by the carrier concentration due to the two reversed processes involved in the electron scattering. In contrast, the phonon relaxation time from EPC has a strong dependence on the carrier concentration, which becomes comparable with that from the intrinsic phonon-phonon scattering in the high carrier concentration region. The electronic transport coefficients are then obtained via integrating the *k*-dependent electron relaxation time predicted by explicit EPC calculations. For the phonon transport, we demonstrate that including the EPC causes significant reduction on the lattice thermal conductivity, especially at higher carrier concentration. As a result, we find a 6% increase of the room temperature *ZT* value at optimized carrier concentration of about $5\times10^{19}$ cm$^{-3}$. More importantly, the EPC gives a 21% reduction of the lattice thermal conductivity at 1200 K, which leads to a remarkable increase of 24% in the *ZT* value at optimized carrier concentration of $3\times10^{20}$ cm$^{-3}$. Our study not only provides a new insight in understanding the thermal conductivity in heavily doped thermoelectric materials, but also emphasizes the importance and necessity of considering EPC in accurately predicting their figure-of-merit.

## Acknowledgements

We thank financial support from the National Natural Science Foundation (Grant Nos. 11574236 and 51772220). The numerical calculations in this paper have been done on the supercomputing system in the Supercomputing Center of Wuhan University.

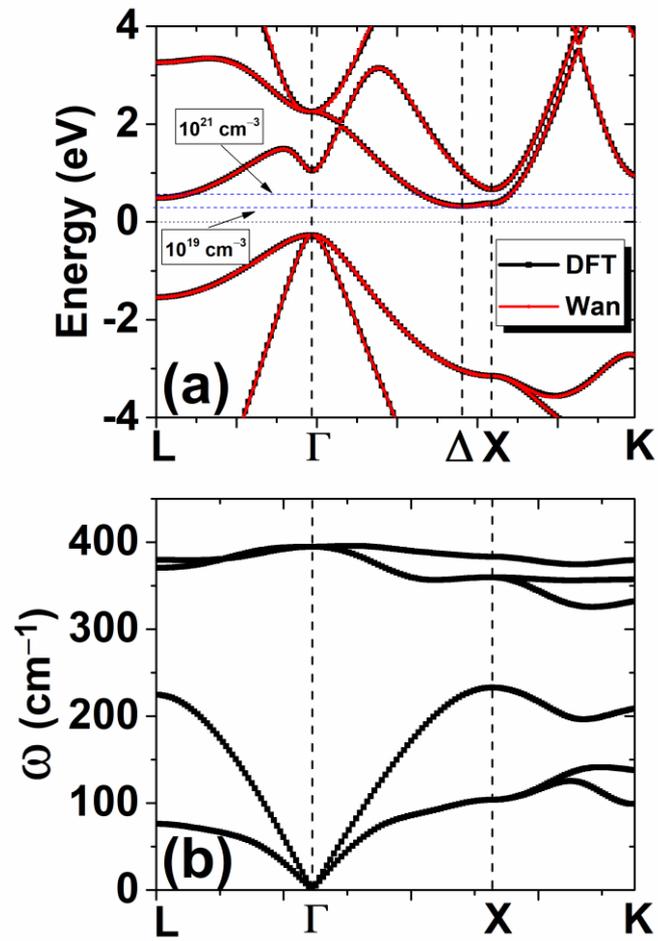

**Figure 1** (a) The electronic band structure of SiGe compound calculated by DFT, as compared with that obtained by Wannier interpolation technique. The two dashed lines indicate the locations of the chemical potential corresponding to the *n*-type carrier concentration of $10^{19}$ and $10^{21}$ cm$^{-3}$. The Fermi level is at 0 eV and is indicated by the dotted line. (b) The phonon spectrum of SiGe compound calculated by DFPT.

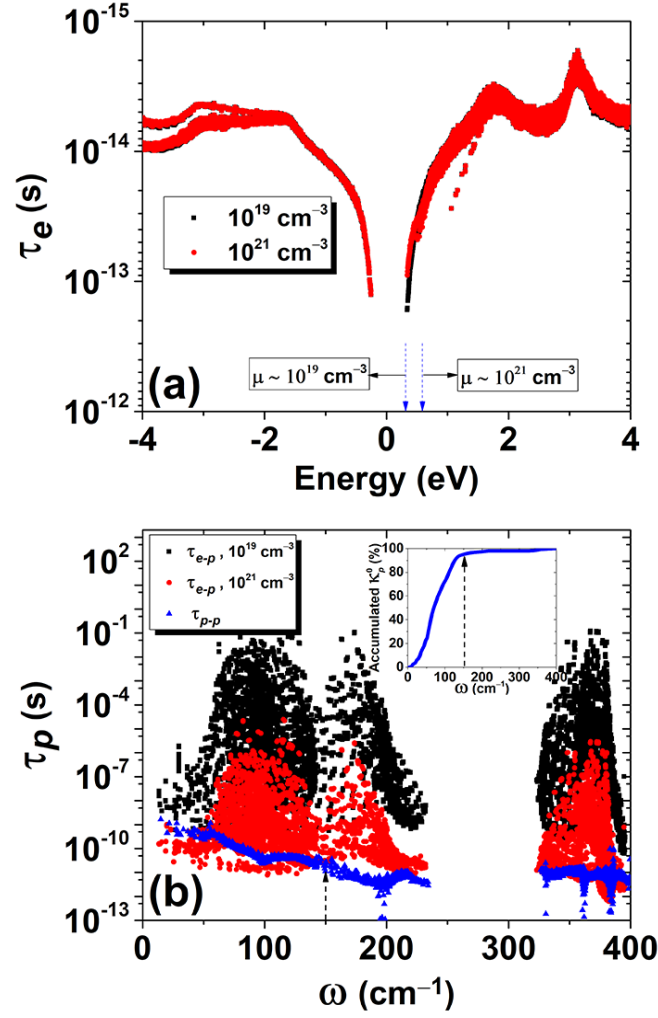

**Figure 2** (a) The electron relaxation time $\tau_e$ of SiGe compound as a function of energy, which is obtained by considering the EPC at two typical carrier concentrations of $10^{19}$ cm$^{-3}$ and $10^{21}$ cm$^{-3}$. The corresponding chemical potentials are also indicated. (b) The phonon relaxation time $\tau_{e-p}$ as a function of frequency, also calculated by considering the EPC at two typical carrier concentrations of $10^{19}$ cm$^{-3}$ and $10^{21}$ cm$^{-3}$. The intrinsic phonon relaxation time $\tau_{p-p}$ is also indicated for comparison. The inset shows the accumulated lattice thermal conductivity as a function of frequency by considering only the intrinsic phonon scattering.

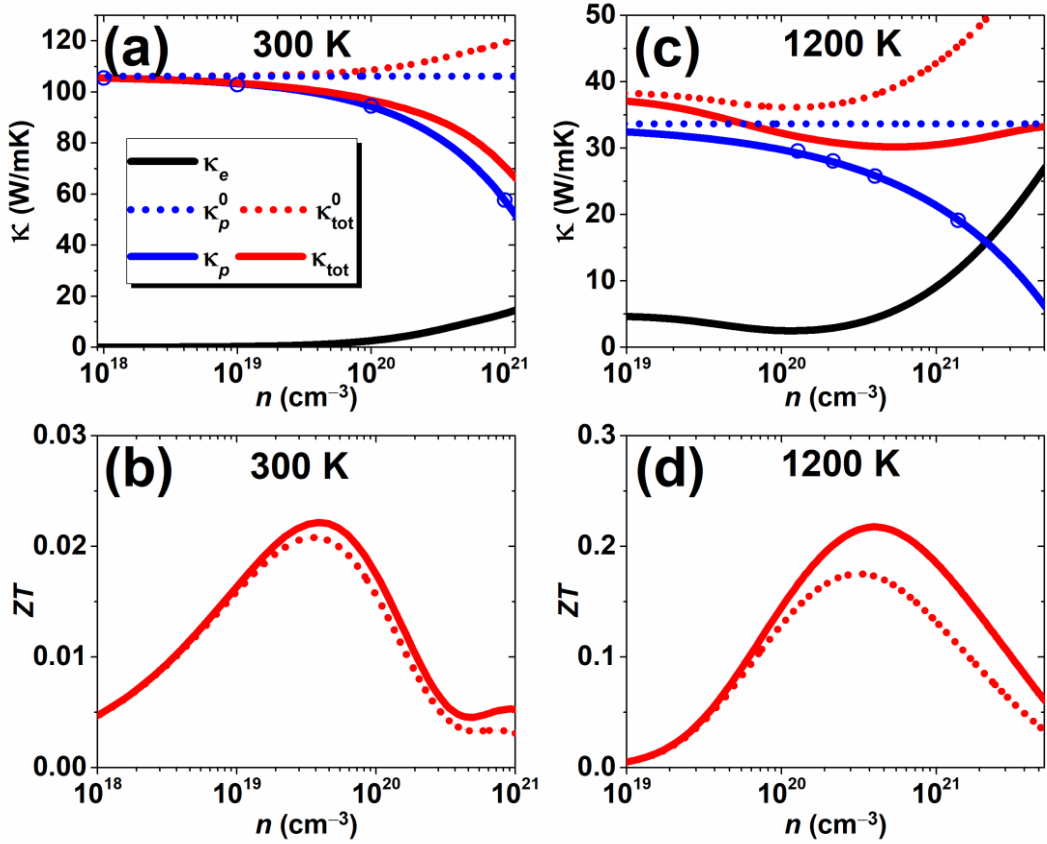

**Figure 3** The thermal conductivity and *ZT* value of SiGe compound as a function of carrier concentration. (a) and (b) correspond to 300 K, while (c) and (d) for 1200 K. The electronic $\kappa_e$ and lattice parts ($\kappa_p$ and $\kappa_p^0$) of thermal conductivity are both shown, and results with and without EPC correction ($\kappa_{tot}$ and $\kappa_{tot}^0$) are compared.